\documentclass[aps,prl,twocolumn,floatfix,showpacs,longbibliography]{revtex4-1}
\usepackage{graphicx}
\usepackage[utf8]{inputenc}
\usepackage[english]{babel}

\usepackage{amssymb}
\usepackage{amsmath}
\usepackage{setspace}
\usepackage{comment}
\usepackage[normalem]{ulem}

\usepackage{soul}
\usepackage[bottom]{footmisc}
\usepackage{color}%
\usepackage[colorlinks=true,linkcolor=blue]{hyperref}%
\hypersetup{allcolors=blue}
\usepackage[normalem]{ulem}
\usepackage{xcolor}
\usepackage{hhline}
\usepackage{tabularx}

\newcommand{\bra}[1]{\ensuremath{\left\langle #1 \right\vert}}
\newcommand{\ket}[1]{\ensuremath{\left\vert #1 \right\rangle}}

\newcommand{\afix}[1]{\textcolor{black}{#1}}
\newcommand{\gfix}[1]{\textcolor{black}{#1}}
\newcommand{\rfix}[1]{\textcolor{black}{#1}}
\begin{document}
\title{AC polarizability and photoionization cross-section measurements in an optical lattice} 
\author{Ryan Cardman}
\email{rcardman@umich.edu}
\author{Xiaoxuan Han}
\altaffiliation[Present address:]{Department of Physics, Taiyuan Normal University, Jinzhong 030619, People’s Republic of China}
\author{Jamie L. MacLennan}
\thanks{R.C., X.H., and J.L.M. made equal contributions to this work.}
\author{Alisher Duspayev}
\author{Georg Raithel}
\affiliation{Department of Physics, University of Michigan, Ann Arbor, MI 48109}
\date{\today}

\begin{abstract}
We use double-resonant two-photon laser spectroscopy to measure the dynamic scalar polarizability of the rubidium 5$D_{3/2}$ level, \gfix{$\alpha^{S}_{5D_{3/2}}$}, at a wavelength of $\lambda=1064$~nm. 
Since $\lambda$ is shorter than the photoionization (PI) limit of the Rb 5$D_{3/2}$ level, \gfix{$\alpha^{S}_{5D_{3/2}}$} depends on
both bound-bound and bound-free transition matrix elements. 
The level also undergoes significant broadening due to PI. The $1064$-nm field is applied in the form of a deep optical lattice ($\sim10^{5}$~photon recoils) generated by an in-vacuum  field-enhancement cavity. In our spectroscopic method, we use known dynamic polarizabilities to eliminate the need to measure the light intensity. Our method yields, in atomic units, $\alpha^{S}_{5D_{3/2}}=-524(17)$, in agreement with estimates. Additionally, we extract the $5D_{3/2}$ photoionization cross section $\sigma$ at $1064$~nm from spectral linewidths; we find $\sigma=44(1)$~Mb.
\end{abstract}
\maketitle

\par For some time, neutral atoms have been trapped by off-resonant optical fields for the purpose of redefining the second with unparalleled precision~\cite{BACON2021}, simulating theoretical models~\cite{LeKien2013, Sanchez.2018, Hu.2019}, and constructing quantum computing protocols~\cite{Saffman.2016, morgado2021}. An atom in an electric field with a frequency far from resonance of an electric-dipole transition undergoes an energy shift due to the AC Stark effect, which is proportional to the \afix{field intensity.} 
\afix{In applications of optical-dipole traps in optical clocks and in spectroscopy, differential AC Stark shifts of the relevant atomic states} must be either very well known or eliminated using carefully determined ``magic'' wavelengths~\cite{SafronovaRbMagic, Flambaum2008, Lundblad2010}.  The AC shifts follow $\Delta W= - \alpha (\omega_L) E_L^2/4$, with laser electric field $E_L$ and angular frequency $\omega_{L}$, and state-dependent dynamic polarizabilities $\alpha$. \gfix{Following the intensity-gradient force on the atomic CM coordinate, $-\nabla_{\bf{R}} \Delta W ({\bf{R}})$, states with positive (negative) $\alpha$ are attracted to (repelled from) locations of high field intensity.
Also, the AC shifts of spectral lines of atomic transitions scale with the difference between the $\alpha$-values of the relevant states.} 

For an atomic state $\ket{n,l,j,m_{j}}$ in a field with polarization unit vector $\hat{\epsilon}$, the dynamic polarizability is
\begin{multline}
    \alpha_{n,l,j,m_{j}}(\omega_{L})=\frac{2}{\hbar}\bigg[\sum_{n',l',j',m'_{j}}|\bra{n',l',j',m'_{j}}\hat{\epsilon}\cdot\hat{\textbf{d}}\ket{n,l,j,m_{j}}|^{2}\\
    \times\frac{\omega^{n'l'j'}_{nlj}}{(\omega^{n'l'j'}_{nlj})^2-\omega^{2}_{L}}
    +\sum_{l',j',m'_{j}}\int_0^\infty d\epsilon'\rho(\epsilon')\\
    \times|\bra{\epsilon',l',j',m'_{j}}\hat{\epsilon}\cdot\hat{\textbf{d}}\ket{n,l,j,m_{j}}|^{2}
     \frac{\omega^{\epsilon'}_{nlj}}{(\omega^{\epsilon'}_{nlj})^2-\omega^{2}_{L}}\bigg],
\label{eq:polarizability}
\end{multline}
where $\hat{\textbf{d}}$ is the electric-dipole moment operator, $\omega^{n'l'j'}_{nlj}=(W_{n'l'j'}-W_{nlj})/\hbar$ for bound state energies $W_{nlj}$ and $W_{n'l'j'}$, and $\omega^{\epsilon'}_{nlj} = (\epsilon'-W_{nlj})/\hbar$ for free-electron-state (FES) energies $\epsilon'$. The quantity $\rho(\epsilon')$ is the density of FESs with energy $\epsilon'$ and is equal to one per unit energy, for energy-normalized states~\cite{friedrichbook}. When the photon energy $\hbar\omega_{L}$ is large enough to photoionize the atom, the second term in Eq.~\ref{eq:polarizability} may become significant because the integrand has a pole at $\omega_L = \omega_{nlj}^{\epsilon'}$. \gfix{In a linearly polarized field,} the $m_{j}$-dependent polarizabilities $\alpha_{n,l,j,m_{j}}(\omega_{L})$ \gfix{depend on} $m_{j}$-independent scalar and tensor polarizabilities, $\alpha^{S}_{n,l,j}(\omega_{L})$ and $\alpha^{T}_{n,l,j}(\omega_{L})$,
\begin{equation}
    \alpha_{n,l,j,m_{j}}(\omega_{L})=\alpha^{S}_{n,l,j}(\omega_{L})+\frac{3m^{2}_{j}-j(j+1)}{j(2j-1)}\alpha^{T}_{n,l,j}(\omega_{L}),
    \label{eq:scalartensor}
\end{equation}
where the second term vanishes for $j<1$. A third term, proportional to the vector polarizability $\alpha^{V}_{n,l,j}(\omega_{L})$, is added to Eq.~\ref{eq:scalartensor} only when the field polarization \gfix{is not linear}.

\par In this study, we measure $\alpha^{S}_{5D_{3/2}}$ for rubidium in an optical lattice of laser wavelength $\lambda = 1064$~nm. Its photon energy $h c / \lambda$ suffices to photoionize $5D_{3/2}$-atoms trapped in the optical lattice, making a case in which the FES-term in Eq.~\ref{eq:polarizability} \gfix{may become significant}. While no theoretical or experimental estimate has been made for the $5D_{3/2}$ dynamic polarizability, the static polarizabilities have been investigated \cite{davydkin1982radiation,kamenski2006jpbelectric,kondratev2008blpipolarizability,snigirev2014prameasurement}, as well as the dynamic polarizability at 778.1~nm for 5$D_{5/2}$ \cite{martin2019prafrequency}. 

The Rb $5D_{3/2}$-levels in a 1064-nm field are also broadened by photoionization (PI), as estimates based on measured Rb $5D_{5/2}$~\cite{duncan2001prameasurement} \rfix{and calculated~\cite{duncan2001prameasurement,Aymar1984} Rb $5D$ PI cross sections} show. 
These estimates exhibit that  the PI-induced level broadening 
at 1064~nm is on the same order of magnitude as the AC Stark shift.
In light of this fact, a comprehensive spectroscopic study of Rb $5D$-states in a 1064-nm optical lattice appears particularly worthwhile, as both the \rfix{ dynamic scalar polarizability} and the PI cross section can be measured \rfix{simultaneously}.
From an applications point of view, the Rb 5$D$ states are appealing to study for several reasons. First, there has been continued interest ranging from earlier decades~\cite{nez1993optcomoptical,touahri1997optcomfrequency,hilico1998epjapmetrological} to recent years \cite{terra2016apbultra,rathod2017sacaccessing,martin2018pracompact} in using the strong and narrow (natural linewidth $<$~1~MHz) two-photon transition 5$S_{1/2}\rightarrow$5$D_J$ as an optical frequency reference, which necessitates precise calculation or cancellation of the relevant light shifts~\cite{gerginov2018twoprap,martin2019frequency}. This two-photon transition's wavelength (778~nm) can be directly generated by a semiconductor laser, or via second-harmonic-generation of a laser at 1556~nm, which falls within the telecommunication band. Furthermore, the 5S$_{1/2} \rightarrow $5D$_{5/2}$ transition is one of the recommended transitions by the Consultative Committee of Length (CCL) for the practical definition of the meter~\cite{quinn2003metpractical}.

\begin{figure}[t]
 \centering
  \includegraphics[width=0.46\textwidth]{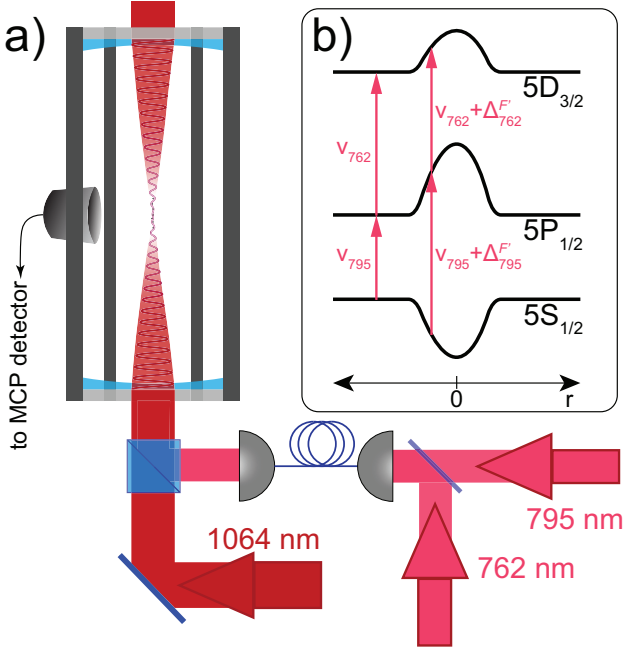}
  \caption{\gfix{(a) Sketch of the experimental apparatus, and (b)} energy level diagram with outlined AC Stark shifts as a function of radial position with respect to the axis of the 1064~nm optical lattice (not to scale).}
  \label{apparatus}
\end{figure}

\par \gfix{Main components of} the apparatus, atomic energy levels, and optical beam geometries are exhibited in Fig.~\ref{apparatus}. We load $^{85}$Rb atoms from a MOT,
\gfix{initially prepared in the $F=3$ ground-state hyperfine level,} into the \rfix{near-perfect} TEM$_{00}$~mode of a near-concentric \rfix{field-enhancement} cavity with a finesse of $\approx 600$ at $\lambda= 1064~$nm~\cite{Chen2014praatomtrapping, Chen2015, MacLennan2019}.
\gfix{After loading,
the lattice} is adiabatically ramped to a maximum depth of $\sim10^5E_{r}$, where $E_{r}=h\times2.076$~kHz is the photon recoil for $^{85}$Rb at 1064~nm. In order to observe the lattice-shifts affecting the D1 line, we pulse a $795$-nm probe laser for a duration of $\sim15~\mu$s and scan it from -36~MHz to $1344$~MHz with respect to the frequency of the $\ket{5S_{1/2},F=3}\rightarrow\ket{5P_{1/2},F'=2}$ transition. A co-propagating 762-nm probe laser is pulsed on for $\sim500$~ns and is independently scanned from 57~MHz to -843~MHz with respect to the lattice-field-free frequency difference between the $\ket{5P_{1/2},F'=2}$ and $\ket{5D_{3/2},F''=3}$ states. Lattice-induced PI of atoms excited into $5D_{3/2}$ yields a count of atoms in this state when the resulting ions are guided onto a micro-channel plate detector (MCP) with an extraction voltage. This procedure yields a two-dimensional map of ion counts as a function of the two probe laser frequencies, as shown in Fig.~\ref{fig:map}(a). 

\begin{figure}
 \begin{center}
 \includegraphics[scale=0.3]{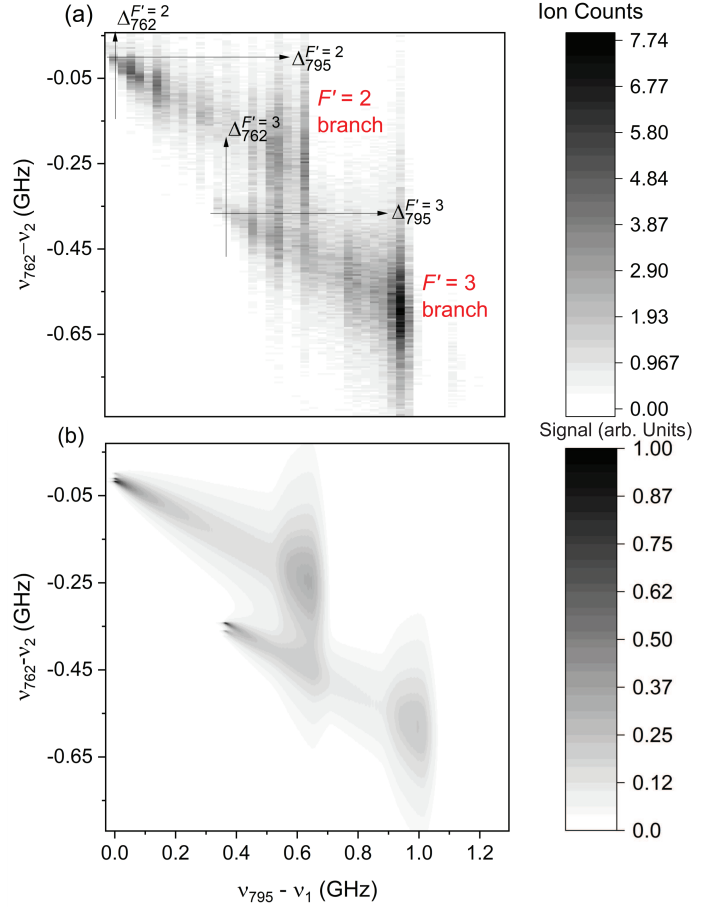}
  \end{center}
  \caption{(a) Results of two-photon spectroscopy as a function of 795-nm and 762-nm probe laser frequencies ($\nu_{795}$ and $\nu_{762}$, respectively). The reference frequencies $\nu_{1}$ and $\nu_{2}$ correspond to splittings between the lattice-field-free $5S_{1/2}, F=3$ and $5P_{1/2}, F'=2$ levels, and between the lattice-field-free $5P_{1/2}, F'=2$ and $5D_{3/2}, F''=3$ levels, respectively. Data are averaged over 30 experimental cycles. (b) Numerical simulation (see Supplement) for input parameters $\alpha^{S}_{5D_{3/2}}=-524$, $\alpha^{T}_{5D_{3/2}}=0$ and $\sigma=40$~Mb.}
  \label{fig:map}
\end{figure}

In Fig.~\ref{fig:map}(a) we display ion counts as a function of the frequencies of both probe lasers. 
The figure exhibits two diagonally-aligned branches of $5D_{3/2}$ atomic signals. The branches correspond to the intermediate hyperfine states $F'=2$ and 3, which present two pathways through which the atoms can be excited into $5D_{3/2}$ via step-wise, double-resonant two-photon excitation.
For the two pathways $F'=2$ and 3, we \gfix{define detunings} $\Delta^{F'}_{795}$ and $\Delta^{F'}_{762}$ of the excitation lasers from their respective lattice-field-free resonances (see Fig.~\ref{apparatus}(b) and axis insets in Fig.~\ref{fig:map}(a)).

The AC shifts evident from the slopes of the signal branches in Fig.~\ref{fig:map} are caused by the dynamic polarizabilities of all relevant atomic states, $5S_{1/2}, F=3$, and 
$5P_{1/2}, F'=2$ or 3, and
$5D_{3/2}, F'= 1$ to $4$. For the ground and intermediate-state polarizabilities at 1064~nm, we use the theoretical values $\alpha^{S}_{5S_{1/2}}=687.3(5)$~\cite{Arora2012} and $\alpha^{S}_{5P_{1/2}}=-1226(18)$~\cite{Neuzner2015} (in atomic units, and applicable to all hyperfine sublevels). The objective of the measurement then is, in principle, to extract the scalar and tensor polarizabilities, $\alpha^{S}_{5D_{3/2}}$ and $\alpha^{T}_{5D_{3/2}}$, from the slopes of the signal branches in Fig.~\ref{fig:map}. From an estimate \gfix{given} below, it is predicted that the magnitude of $\alpha^{T}_{5D_{3/2}}$ is at or below the level of uncertainty of the experimental method. 
In the following, we therefore restrict our analysis to the case $\alpha^{T}_{5D_{3/2}}=0$.

The two signal branches in Fig.~\ref{fig:map} yield two measurements for the slopes, $d\Delta^{F'}_{762}/d\Delta^{F'}_{795}$, associated with the two intermediate states $5P_{1/2}, F'=2$ and $5P_{1/2}, F'=3$. The scalar polarizability $\alpha^{S}_{5D_{3/2}}$ then follows from
\begin{equation}
    \alpha^{S}_{5D_{3/2}}(F')=\alpha^{S}_{5_{P_{1/2}}}-\frac{d\Delta^{F'}_{762}}{d\Delta^{F'}_{795}}(\alpha^{S}_{5S_{1/2}}-\alpha^{S}_{5P_{1/2}}) \quad.
\label{eq:slope}   
\end{equation}
\noindent \gfix{This measurement method is similar to a method used in~\cite{Chen2015}}.
Because Fig.~\ref{fig:map}(a) yields two readings for the slope $d\Delta^{F'}_{762}/d\Delta^{F'}_{795}$, we obtain two measurements for $\alpha^{S}_{5D_{3/2}}$ that correspond with the intermediate hyperfine pathways $F'=2$ and $F'=3$. \gfix{The method} is self-calibrating in the sense that a direct measurement of the lattice intensity in the atomic sampling region is unnecessary.

\begin{table}[htbp]
\caption{\label{tab:table1} Summary of quantities used to extract $\alpha^{S}_{5D_{3/2}}$, as well as a summary of results \gfix{(in atomic units)}.}
\begin{ruledtabular}
\begin{tabular}{c|c|c}
    Quantity & Value & Source\\
    \hline
    $\alpha^{S}_{5S_{1/2}}$ & 687.3(5) & \cite{Arora2012} \\ 
    $\alpha^{S}_{5P_{1/2}}$ & -1226(18) & \cite{Neuzner2015} \\
    $d\Delta^{F'=2}_{762}/d\Delta^{F'=2}_{795}$ & -0.36(1) & Experimental data, this work \\
    $d\Delta^{F'=3}_{762}/d\Delta^{F'=3}_{795}$ & -0.371(6) & Experimental data, this work\\
    \hline
    $\alpha^{S}_{5D_{3/2}}(F'=2)$ & -537(27) & Eq.~\ref{eq:slope}\\
    $\alpha^{S}_{5D_{3/2}}(F'=3)$ & -516(22) & Eq.~\ref{eq:slope}\\
    $\alpha^{S}_{5D_{3/2}}$ & -524(17) & Weighted average\\
\end{tabular}
\end{ruledtabular}
\end{table}

The experimental data are acquired by scanning the detuning of the 762-nm laser, $\nu_{762}-\nu_2$, for a set of values for the detuning $\nu_{795}-\nu_1$ of the 795-nm laser.
A \rfix{vertical} slice of the map in Fig.~\ref{fig:map}(a) at a value of $\nu_{795}-\nu_{1}=0.984$~GHz is provided in \rfix{the inset of} Fig.~\ref{fig:crosssection}(a) as an example. We find that the spectral lines along each frequency setting of the 795-nm laser follow near-perfect Lorentzian shapes, down to the noise level several linewidths away from the line centers. 
This observation serves as experimental \gfix{evidence} that the line profiles are the result of a level decay mechanism, which in our case is the PI of the $5D_{3/2}$ levels. For each of the individual spectra at fixed  $\nu_{795}-\nu_1$, we obtain the line centers and linewidths using Lorentzian fits. Depending on whether the signal has contributions from both signal branches  $F'=2$ and $F'=3$ in Fig.~\ref{fig:map} or \gfix{from} just one, we employ double- or single-Lorentzian fits, respectively. The line centers yield two sets of data points $\Delta^{F'}_{762} (\Delta^{F'}_{795})$, for $F'=2$ and $3$. \rfix{We present the sets for the $F'=3$ branch in Fig.~\ref{fig:crosssection}(b).} The slopes $d\Delta^{F'}_{762}/d\Delta^{F'}_{795}$ from respective weighted linear fits yield a pair of measured values of $\alpha^{S}_{5D_{3/2}}$ via Eq.~\ref{eq:slope}. \gfix{Essential inputs from other sources and our results}  are summarized in Table~\ref{tab:table1}. The uncertainties in the scalar polarizability measurements are dominated by the statistical errors of the linear fits and the given uncertainty in $\alpha^{S}_{5P_{1/2}}$. Systematic effects from stray electromagnetic fields, atomic collisions, and laser frequency linearity are negligible. The weighted average over the pair of measured dynamic scalar polarizabilities is $\alpha^{S}_{5D_{3/2}}=-524(17)$ \gfix{atomic units}.

\gfix{Expanding upon previous results from other works, theoretical estimates for $\alpha^{S}_{5D_{3/2}}$ and $\alpha^{T}_{5D_{3/2}}$ can be obtained}.
In Refs.~\cite{safronova2004prarelativistic,safronova2011pracritically}, the listing of either the dipole matrix elements themselves or the contributions of each transition to other polarizabilities allow the calculation of the equivalent contribution for the dynamic polarizability at an arbitrary frequency, using the first term in Eq.~\ref{eq:polarizability}. For the $(n\geq 7)F_{5/2}$ and $(n\geq 9)P_{J}$ contributions, we \gfix{use dipole} matrix elements from an online atom calculator \cite{vsibalic2017arc} or from our own calculations~\cite{Reinhard2007}. In this way, we estimate \gfix{the bound-bound contributions to the dynamic polarizability (discrete sum in Eq.~\ref{eq:polarizability})}.
To estimate the continuum integral in Eq.~\ref{eq:polarizability}, we compute bound-free matrix elements using model potentials given in~\cite{Marinescu} on a dense \gfix{FES} energy grid (\gfix{spacing} $\sim h \times 40$~GHz) over a range from $\epsilon'=0$ up to $\sim20$~eV, yielding continuum contributions of \gfix{18 for} $\alpha^{S}_{5D_{3/2}}$ and -3 for $\alpha^{T}_{5D_{3/2}}$. \gfix{The core contribution to $\alpha^{S}_{5D_{3/2}}$, not shown in Eq.~1, is estimated at a value of 9, provided in \cite{martin2019prafrequency} for DC and for 778~nm}. Our resulting estimates for the net scalar and tensor polarizabilities, \gfix{including all mentioned contributions,} \gfix{are $\alpha^{S}_{5D_{3/2}}=-516$} and $\alpha^{T}_{5D_{3/2}}=21$. 
From our simulation described in the Supplementary Material, we estimate that the presence of a tensor polarizability of 21 would cause a positive shift of $\sim12$ in the deduced scalar polarizability, when using Eq.~\ref{eq:slope}. Therefore, we conclude that the estimated systematic uncertainty \gfix{in $\alpha^{S}_{5D_{3/2}}$} related to the tensor polarizability lies within the uncertainty of $\pm 17$ for the measured $\alpha^{S}_{5D_{3/2}}$ reported in Table~\ref{tab:table1}.


\begin{figure}
 \begin{centering}
 \includegraphics[scale=0.6]{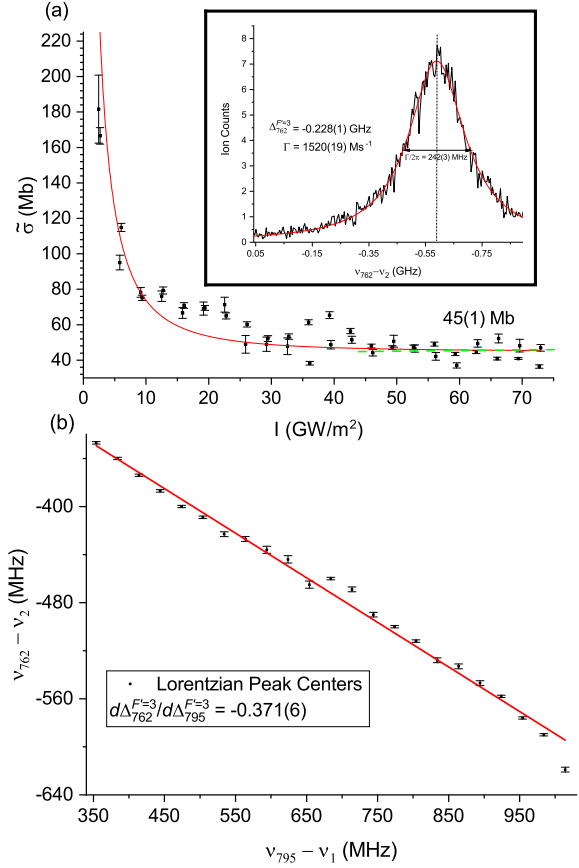}
  \end{centering}
  \caption{\rfix{(a) $\tilde{\sigma}$ vs  $I$ at the atoms' locations. The data and a fit result depicted in red yield a $5D_{3/2}$ photoionization cross section of $\sigma = 45(1)$~Mb. The inset shows a typical spectral peak (a single, vertical slice from Fig.~\ref{fig:map}(a))  characterized by the 762-nm laser detuning at a fixed 795-nm frequency setting ($\nu_{795}-\nu_{1}=0.984$~GHz), along with $\Gamma$ and $\Delta^{F'=3}_{762}$ obtained from its Lorentzian fitting parameters. We average ion counts over 30 experimental cycles. (b) Lorentzian peak centers in
  $\nu_{762}-\nu_{2}$ and their uncertainties versus $\nu_{795}-\nu_{1}$ for the $F'=3$ signal branch in Fig.~2~(a).} The slope of the weighted linear fit to these data \gfix{is $d\Delta^{F'}_{762}/d\Delta^{F'}_{795}=-0.371(6)$}. We repeat this procedure for the $F'=2$ branch. }
  \label{fig:crosssection}
\end{figure}

In the following, we discuss a measurement of the $5D_{3/2}$ PI cross section. For a given detuning $\Delta^{F'}_{795}$, the 1064-nm light intensity $I$ at the locations of atoms contributing to the peak in the ion signal in the $F'$ branch is given by $I=2 h \Delta^{F'}_{795} c \epsilon_0 /(\alpha^S_{5S_{1/2}}-\alpha^S_{5P_{1/2}})$,
with an uncertainty arising from the polarizabilities and the natural linewidth of the rubidium D1 line (5.75~MHz~\cite{SteckRb85}). The linewidth of ion spectra for fixed $\Delta^{F'}_{795}$ and scanned $\Delta^{F'}_{762}$, $\Gamma/2\pi$, obtained from Lorentzian fits as shown in Fig.~\ref{fig:crosssection}(a), then provides an upper limit of the PI decay rate $\Gamma_{PI}$ at intensity $I$.
Using 
\begin{equation}
    \Gamma_{PI}=\frac{\sigma I}{\hbar \omega_{L}},
\label{eq:decayrate}
\end{equation}
for each measured \rfix{decay rate} $\Gamma$(I) we obtain an upper bound $\tilde{\sigma} (I) = \hbar \omega_L \Gamma(I) / I$ for the Rb~$5D_{3/2}$ PI cross section $\sigma$. For atoms located near the bottoms of the lattice wells, corresponding to the largest $\Delta^{F'}_{795}$ and the largest intensities $I$, the broadening is near-exclusively given by PI. In contrast, at the lowest $\Delta^{F'}_{795}$ and intensities $I$ other mechanisms, such as the excited-state hyperfine coupling and residual off-resonant two-photon signals, are principal.
As a result, $\tilde{\sigma} (I) \approx \sigma$ at large $I$ and  $\tilde{\sigma} (I) >> \sigma$ at low $I$. This effect is exhibited in Fig.~\ref{fig:crosssection}(a), where the quantity $\tilde{\sigma}$ converges to $\sigma$ (and $\Gamma$ to $\Gamma_{PI}$) at the high-intensity end of the $I$-axis.
\par Quantitatively, we obtain $\sigma$ from the data in Fig.~\ref{fig:crosssection}(a) using two methods. In Method A, we take the arithmetic average of $\tilde{\sigma}$ in the asymptotic region $I\gtrsim45$~GW/m$^{2}$ and utilize the standard error of the mean as its uncertainty. Method A yields $\sigma=45(1)$~Mb. In Method B, we apply a fit function $\tilde{\sigma}=\sqrt{\sigma^{2}+\gamma^{2}/I^{2}}$, \gfix{where $\gamma$ accounts for} broadening mechanisms other than PI. This fit also gives $\sigma=45(1)$~Mb. Combining the two methods, we have $\sigma=45(1)$~Mb. 
Our simulations, \gfix{explained in the Supplement}, show that the measurement method likely overestimates the PI cross section of an isotropic atom sample by $\approx 1.7\%$, leading to \gfix{our} slightly corrected final result of $\sigma=44(1)$~Mb. Based on the good qualitative agreement of measured and simulated strength ratios between the $F'=2$ and $F'=3$ signal bars in Fig.~\ref{fig:map}, where the atom sample in the simulation is isotropic, we do not believe that optical pumping causes a significant deviation of our measured PI cross section from what it would be under perfectly isotropic conditions.

In a calculation based on fine-structure-less model potentials from~\cite{Marinescu}, we have found a total shell-averaged PI cross section of 32.4~Mb at 1064~nm and 43.7~Mb at the PI threshold wavelength, which is close to calculations in ~\cite{duncan2001prameasurement} and~\cite{Aymar1984}. Trap-loss measurements in~\cite{duncan2001prameasurement} for Rb~$5D_{5/2}$ gave a result of 18~Mb at 1064~nm and an estimate of 25~Mb at threshold. The discrepancies between the results for the Rb $5D$ PI cross sections await a future explanation.

\par In summary, we have spectroscopically measured the dynamic scalar polarizability of the rubidium $5D_{3/2}$ state in a 1064~nm optical lattice using two probe lasers at 795~nm and 762~nm. We report $\alpha^S_{5D_{3/2}}=-524(17)$ and estimate $\vert \alpha^T_{5D_{3/2}}\vert  << \vert \alpha^S_{5D_{3/2}} \vert $.
The observed PI-induced line broadening has yielded a PI cross section of $\sigma=44(1)$~Mb. Future experimental directions involve the measurement of $\alpha^{S}_{5P_{1/2}}$ with a Rydberg excitation field probing the 1064-nm lattice shifts of the intermediate $5P_{1/2}$ state and \gfix{auxiliary} Rydberg levels~\cite{Chen2015}.
Our measurement will also aid in the preparation of recently-predicted Rydberg-atom-ion molecules~\cite{duspayev2021} and of novel high-angular-momentum Rydberg states in a deep, 1064-nm Rydberg-atom optical lattice~\cite{Younge2010, Cardman2021} using step-wise excitation via an intermediate Rb $5D$ state. It would be also desirable to obtain theoretical estimates of $\alpha^{S}_{5D_{3/2}}$ and improved theoretical values for $\sigma$ to compare with experimental results.

\maketitle
\section*{ACKNOWLEDGMENTS}
This work was supported by NSF grant No. PHY-1806809. X.H. acknowledges support from the program of China Scholarships Council (No. 201808140193). We would like to thank Yun-Jhih Chen for initial experimental work.

\bibliographystyle{apsrev4-1}
\bibliography{pol_exp.bib}

\end{document}


\section{Supplementary Material}
\subsection{Pound-Drever-Hall (PDH) Locking of Field-Enhancement Cavity}

\gfix{The $1064$-nm optical lattice is derived from the TEM$_{00}$ mode of an in-vacuum, optical-field-enhancement cavity (finesse of 600)~\cite{Chen2014praatomtrapping}. This near-concentric cavity features resonant peaks corresponding to TEM modes that are a function of the cavity length and laser frequency. Situated on a vibrationally isolated laser table, the in-vacuum cavity is still susceptible to mechanical vibrations that can disrupt the resonator's length. The mechanical oscillations occur mostly in the $\lesssim 1$~kHz-regime and are compensated by a piezo-electric transducer (PZT) that controls the cavity length (``slow lock"). The slow lock also compensates for thermal drifts of the 1064-nm lattice laser, which has a short-term linewidth of about 100~kHz. 
High-frequency noise of the laser frequency 
is eliminated by frequency modulation of an acousto-optical modulator (AOM) using 
a "fast lock" circuit. Both the slow and fast locks employ an error signal obtained with the Pound-Drever-Hall (PDH) method~\cite{Drever1983,Black2001}. In our PDH implementation, we send the $-2^{\text{nd}}$ order of a
double-pass AOM, driven at $\simeq40$~MHz, through an EOM driven 
at $\simeq20$~MHz. A portion of the laser light is sampled by PD1 and sent as a reference signal to a normalization chip (NORM). The reflected light from the cavity is photo-detected by PD2. We mix the PD2 signal with the EOM drive ($\simeq20$~MHz) in a phase-sensitive PDH detector, and send the mixed-down PDH error signal through a low-pass filter (LP1). The low-passed PDH error signal is employed in both the slow and the fast locks of the TEM$_{00}$ cavity mode.} 

\begin{figure}[t!]
 \centering
  \includegraphics[scale=0.5]{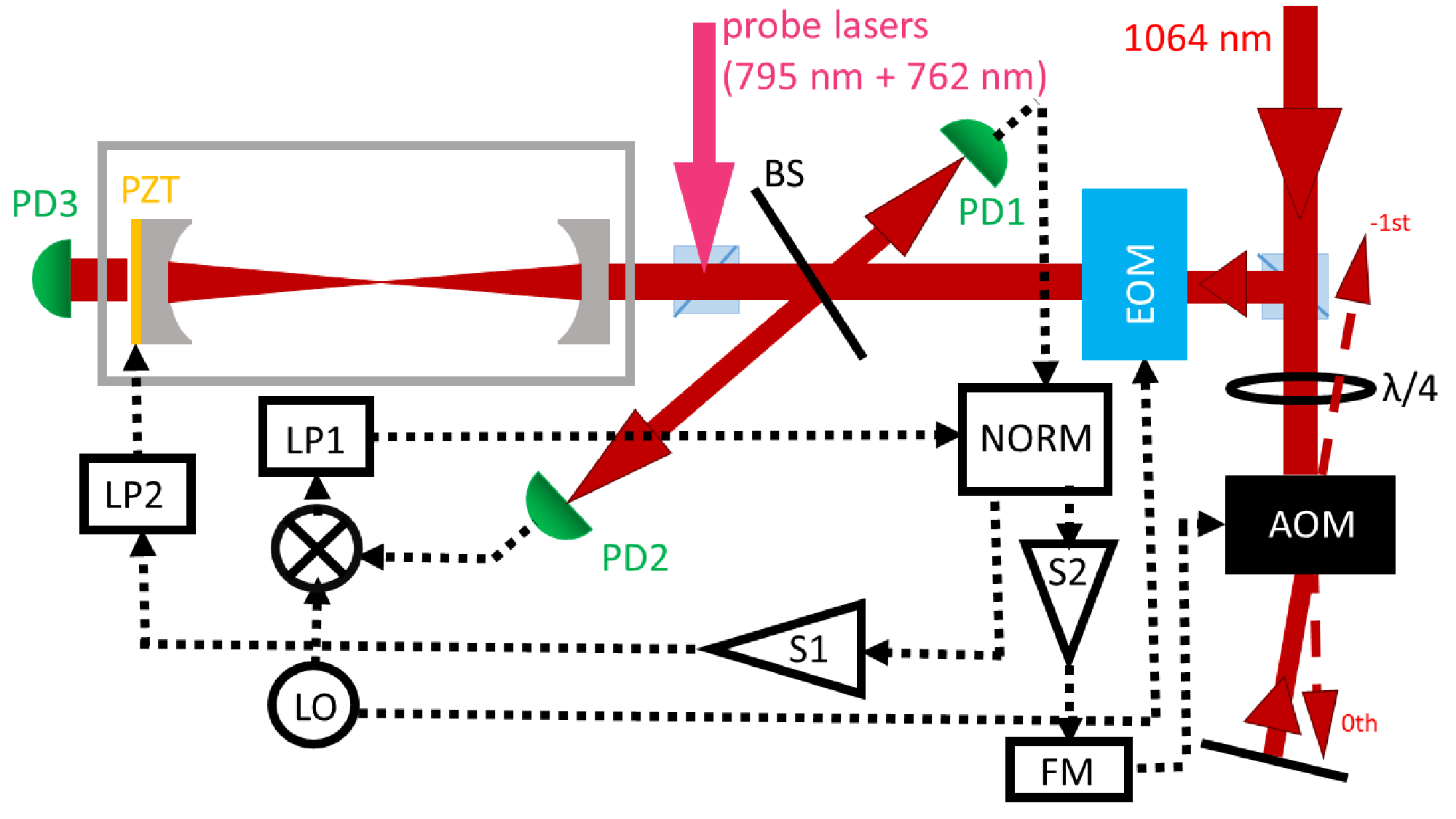}
  \caption{Slow- and fast-locking mechanism for maintaining a TEM$_{00}$ lattice mode. \gfix{The labels are: ``PD1",``PD2",``PD3" photodiodes} and transimpedance amplifiers; ``BS" beam sampler (96:4 transmission-to-reflection ratio); ``PZT" piezo-electric transducer; ``EOM" electro-optical phase modulator; ``AOM" acousto-optical modulator;``$\lambda/4$" quarter-wave plate; ``NORM" normalization circuit (see text for details); ``LO" oscillator driving EOM ($\simeq20$~MHz); ``LP1",``LP2" low-pass filters; ``S1",``S2" servo amplifiers; ``FM" frequency-modulation signal for the RF drive of the AOM.  }
  \label{fig:lock}
\end{figure}

\gfix{The NORM circuit, which outputs the quotient of the low-passed PDH error signal after LP1 and the lattice-intensity reference signal from PD1, is necessary because the AOM is amplitude-modulated for adiabatic ramping of the lattice intensity to a large maximum value of $\sim10^{5}E_{r}$ within the cavity (recoil energy $E_r=2.076~$kHz). The resultant strong (synchronous) variation of the PD1 and the low-passed PDH error signals requires a normalization.  
The normalized PDH error signal from the NORM circuit is split and sent to two separate servo amplifiers, S1 and S2. The output of S1 effects the slow lock, which acts on a ring-shaped PZT attached to a cavity mirror located within the vacuum chamber. The output of S2 acts on the frequency-modulation input of the $\simeq40$-MHz driver of the AOM, effecting the fast lock.}

\subsection{Phased-Lock Loops for Two-Photon Laser Spectroscopy}

\begin{figure}[t!]
 \centering
  \includegraphics[scale=0.50]{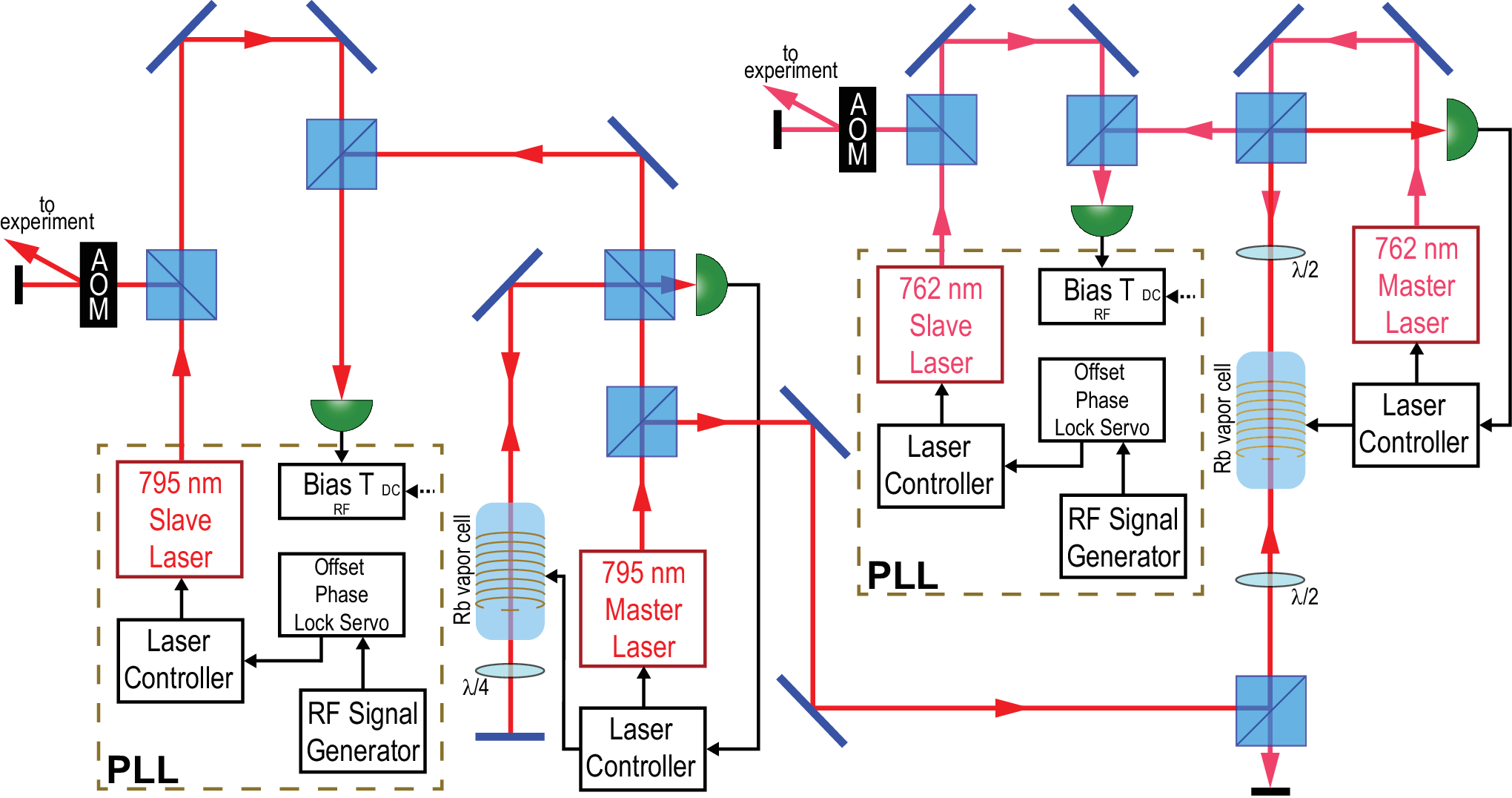}
  \caption{Phase-locked loops for frequency control of excitation lasers.}
  \label{fig:plls}
\end{figure}

Our experiment relies on linear scanning of both the 795-nm and 762-nm probe lasers \gfix{over wide spectral ranges. Linearity of the laser scans is crucial for measurements of $\frac{d\Delta_{762}^{F'}}{d\Delta_{795}^{F'}}$ for determining the $5D_{3/2}$ dynamic polarizability, and for measurement of the photoionization cross section. Usually, linear current or linear voltage control of the laser current or the piezo-electric (PZT) 
in an external-cavity diode laser does not result in accurate, linear scans over ranges of several GHz.} For this reason, we lock the probe lasers (which we denote as ``slave lasers" in this section) relative to fixed-frequency master lasers that are frequency-stabilized to atomic references.

The complete schematic for our optical phase-locked loops (PLLs) is depicted in Fig.~\ref{fig:plls}. The master cateye-etalon diode laser (CEDL) operating at $795$~nm is peak-locked to the $^{87}$Rb $F=2\rightarrow F'=1$ hyperfine transition of the D1 line with saturation-absorption spectroscopy (SAS) in a $3$~cm long vapor cell~\cite{SteckRb85}. Peak locking is achieved through the creation of an error signal by means of dithering the current sent through coils wrapped around a rubidium vapor cell at a frequency of $250$~kHz. The resulting SAS is modulated through the Zeeman effect at $250$~kHz. A portion of the 795-nm master CEDL is sent through a separate vapor cell and counterpropagates with a beam from the 762-nm master CEDL in order to provide electromagnetically induced transparency (EIT) spectroscopy of the $^{87}$Rb $5P_{1/2}\rightarrow5D_{3/2}$ line upon photodetection of the 795-nm beam. In this spectrum, the hyperfine transitions of the $5D_{3/2}$-line are resolvable. We stabilize the 762-nm, master CEDL to the $^{87}$Rb $F'=1\rightarrow F''=2$ hyperfine transition of the $5P_{1/2}\rightarrow5D_{3/2}$ EIT line with a peak lock similar to that of the 795-nm laser. The lowest achievable linewidths for both stabilized master lasers are a few $100$~kHz. 

\gfix{Once both master lasers are-frequency stabilized, we combine them with their corresponding slave lasers on fast photodiodes (several GHz bandwidth). The resulting beat signals, picked off with a bias-T, are demodulated with RF signal generators and sent to high-bandwidth $\sim20$~MHz offset-phase-lock servos that control the laser currents of the 795-nm and 762-nm slave lasers. The RF signal generators used for the demodulation have a frequency stability of $<$~1~kHz. By scanning the frequency of the RF signal generators while the slave lasers are phase-locked, we are able to perform GHz-long linear scans of the two probe lasers with sub-MHz resolution.}   

\subsection{Simulation}

\par We simulate the double-resonant two-photon laser spectrum by considering stationary atoms within a one-dimensional 1064-nm optical lattice with a maximum 5$S_{1/2}$ AC shift of -254~MHz, as determined from experimental spectra. The distribution of atoms as a function of the 5$S_{1/2}$ AC shift, $P(\Delta_{5S_{1/2}})$, is spectroscopically measured and entered into the simulation. The atomic quantization axis is along the lattice propagation direction $\bf{\hat{z}}$. The 1064-nm, $795$-nm and $762$-nm lasers are linearly polarized along 
$\bf{\hat{y}}$, $\bf{\hat{x}}$ and $\bf{\hat{x}}$, respectively, as in the experiment. We represent the Hamiltonian in the basis 
$\{ \vert 5S_{1/2}, m_j, m_i \rangle,  \vert 5P_{1/2}, m_j, m_i \rangle,
\vert 5D_{3/2}, m_j, m_i \rangle \} $  for $^{85}$Rb, with all 48 electronic and nuclear magnetic sublevels of the system included. 

We include the hyperfine interaction~\cite{SteckRb85} and the light-shift interaction, given by Eq.~1 in the main text multiplied by $-E_L^2/4$, with local lattice electric-field amplitude $E_L$.
For the hyperfine $A$ constants of $5S_{1/2}$ and $5P_{1/2}$ 
we use data from~\cite{SteckRb85}, and for the hyperfine $A$ and $B$ constants of $5D_{3/2}$ we use values from~\cite{nez1993optcomoptical}. For the known AC polarizabilities we enter 
$\alpha^{S}_{5S_{1/2}} = 687.3$~\cite{Arora2012} and $\alpha^{S}_{5P_{1/2}}= -1226$~\cite{Neuzner2015}, in atomic units. For $\alpha^{S}_{5D_{3/2}}$
and $\alpha^{T}_{5D_{3/2}}$ we may enter, for instance, our experimental values of $-524$ and 0, respectively. 

The full Hamiltonian is diagonalized for 1000 equidistant intensity steps ranging from 0 to the intensity value at the anti-nodes of the lattice. This yields three blocks of lattice-mixed eigenstates, for the $5S_{1/2}$, $5P_{1/2}$ and $5D_{3/2}$ levels, respectively. The atoms are assumed to be initially equally distributed over the $F=3$-like states of the $5S_{1/2}$ ground manifold, with an overall weighting over the 1000 intensity steps according to the measured $P(\Delta_{5S_{1/2}}$). 

The photoionization (PI) rate of the 20 lattice-mixed $5D_{3/2}$ states follows from shell-averaged PI cross section, $\sigma$, which is an input parameter that we may set, for instance, at $40~$Mb. The distribution of the lattice-mixed states over the electronic magnetic sublevels then yields state-dependent PI cross sections, $\sigma_k$, that depend somewhat on the lattice-mixed $5D_{3/2}$ states, labeled $k=1, ..., 20$. The state-dependent PI rates then are $\Gamma_{PI, k} = \sigma_k I/ (\hbar \omega_L)$, with local light intensity $I$ and lattice angular frequency $\omega_L$.

Since the intensities of the excitation beams are near or below the respective saturation intensities, and since the upper (762-nm) transition is heavily broadened by PI of the $5D_{3/2}$ levels, the signals can be approximated by two-dimensional spectral profile functions $S(\Delta_{795}, \Delta_{762})$ that are mildly saturated Lorentzians
with a width near 6~MHz along $\Delta_{795}$, and PI-broadened Lorentzians with intensity- and state-dependent frequency linewidths $\Gamma_{PI, k}/(2 \pi)$. The entire spectrum is then assembled by taking a weighted sum of profile functions centered at the detunings of the lattice-shifted lower and upper transitions along the $\Delta_{795}$- and $\Delta_{762}$-directions, respectively, and weighting factors given by the product of the line strengths of the lower and upper transitions, which are computed for the given probe light polarizations. An underlying assumption here is that every atom excited into a $5D_{3/2}$ level is photoionized. This assumption is very well satisfied due to the fact that our PI rates range up into the $10^9~$s$^{-1}$ range. The two-dimensional spectra obtained by this procedure are then finally summed over the
1000 sampled intensity values, which
a weighting function that follows from the measured lower-transition AC-shift profile $P(\Delta_{5S_{1/2}})$ entered into the simulation. 

The simulations, an example of which is shown in Fig.~2(b) of the main text, qualitatively reproduce the measurements very well. The qualitative agreement in signal strengths of the F'=2 and F'=3 branches is an indicator that optical pumping plays no or only a small role in the measurement. Hence, our measured $\sigma$-value is expected be close to what it would be for an ideally isotropic atom sample. Further, the simulated spectral lines along $\Delta_{762}$ can be fitted in a way analogous to the processing of the experimental data in order to \gfix{find line centers and linewidths from the simulation, and to deduce values for the polarizability $\alpha^{S}_{5D_{3/2}}$ and the PI cross section $\sigma$. For simulation input values $\alpha^{S}_{5D_{3/2}}=-524$ and  $\alpha^{T}_{5D_{3/2}}=0$, the value of $\alpha^{S}_{5D_{3/2}}$ deduced from Lorentzian fits of the simulated spectra and from Eq.~3 in the main text is found to be $\alpha^{S}_{5D_{3/2}}= -523$. Further,} the deduced values of $\sigma$ exceed the input value for $\sigma$ by only about 1.7$\%$. \gfix{Hence,} the simulations essentially validate the experimentally applied procedure. In light of the simulation result, we apply a 1.7$\%$ reduction to the $\sigma$-value deduced from the experimental spectra to arrive at our final experimental estimate of $\sigma=44(1)$~Mb. 

\bibliographystyle{apsrev4-1}
\bibliography{supp.bib}